\documentclass[%
 reprint,
 aps,
 prl,nofootinbib,
 superscriptaddress
]{revtex4-2}
\usepackage[hidelinks]{hyperref}
\usepackage{amsmath}
\usepackage{amssymb}
\usepackage{cleveref}

\usepackage{silence}
\usepackage{dcolumn}
\usepackage{bm}

\usepackage{balance}
\usepackage{siunitx}
\usepackage{float}
\usepackage{graphicx}
\usepackage{breqn}
\usepackage{mathtools}
\usepackage{listings}
\usepackage{comment}
\usepackage[normalem]{ulem}

\usepackage{xcolor}

\usepackage{placeins}

\begin{document}

\crefname{figure}{Fig.}{Figs.}
\Crefname{figure}{Figure}{Figures}
\crefname{equation}{Eq.}{Eqs.}
\Crefname{equation}{Equation}{Equations}
\crefname{section}{Sec.}{Secs.}
\Crefname{section}{Section}{Sections}
\creflabelformat{equation}{#2\textup{#1}#3}
\renewcommand{\crefrangeconjunction}{--}

\newcommand{\reftexts}{Ref.~}
\newcommand{\lips}[1]{\tilde{d} #1 \;}
\newcommand{\vecIV}[1]{#1} %
\newcommand{\vecIII}[1]{\vec{#1}} %
\newcommand{\vecII}[1]{\boldsymbol{#1}} %

\makeatletter

\newcommand{\coll}[2]{%
	$\coll@process{#1} + \coll@process{#2}$%
}

\newcommand{\collThree}[3]{%
	$\coll@process{#1} / \coll@process{#2} + \coll@process{#3}$%
}

\newcommand{\collFour}[4]{%
  $\coll@process{#1} / \coll@process{#2} / \coll@process{#3} + \coll@process{#4}$%
}

\newcommand{\cole}[1]{%
	{#1}%
}

\newcommand{\coll@process}[1]{%
  \ifx#1A%
    #1%
  \else\ifx#1B%
    #1%
  \else\ifx#1x%
    #1%
  \else\ifx#1p%
    #1%
  \else\ifx#1d%
    #1%
  \else
    \coll@checkHeThree{#1}%
  \fi\fi\fi\fi\fi%
}

\newcommand{\coll@checkHeThree}[1]{%
  \ifnum\pdfstrcmp{#1}{He3}=0 %
    {}^3\mathrm{He}%
  \else
    \mathrm{#1}%
  \fi
}

\newcommand{\aaa}{$A + A$}

\newcommand{\pp}{$p + p$}
\newcommand{\pbpb}{$\mathrm{Pb} + \mathrm{Pb}$}
\newcommand{\pb}{$\mathrm{Pb}$}
\newcommand{\auau}{$\mathrm{Au} + \mathrm{Au}$}
\newcommand{\au}{$\mathrm{Au}$}
\newcommand{\oo}{$\mathrm{O} + \mathrm{O}$}
\newcommand{\ox}{$\mathrm{O}$}
\newcommand{\nene}{$\mathrm{Ne} + \mathrm{Ne}$}
\newcommand{\neon}{$\mathrm{Ne}$}
\newcommand{\arar}{$\mathrm{Ar} + \mathrm{Ar}$}
\newcommand{\ar}{$\mathrm{Ar}$}
\newcommand{\xexe}{$\mathrm{Xe} + \mathrm{Xe}$}
\newcommand{\xe}{$\mathrm{Xe}$}
\newcommand{\li}{\ensuremath{{}^6\mathrm{Li}}}
\newcommand{\lili}{\ensuremath{{}^6\mathrm{Li} + {}^6\mathrm{Li}}}
\newcommand{\btbt}{${}^{10}\mathrm{B} + {}^{10}\mathrm{B}$}
\newcommand{\bt}{${}^{10}\mathrm{B}$}

\newcommand{\pa}{$p + A$}
\newcommand{\da}{$d + A$}
\newcommand{\pdha}{$p / d / {}^3 \mathrm{He} + A$}
\newcommand{\pdhau}{$p / d / {}^3 \mathrm{He} + \mathrm{Au}$}

\newcommand{\ppb}{\ensuremath{p + \mathrm{Pb}}}
\newcommand{\po}{$p + \mathrm{O}$}
\newcommand{\pau}{$p + \mathrm{Au}$}
\newcommand{\dau}{$d + \mathrm{Au}$}
\newcommand{\dpb}{$d + \mathrm{Pb}$}

\newcommand{\hea}[1]{${}^{#1}\mathrm{He} + A$}

\newcommand{\heau}[1]{${}^{#1}\mathrm{He} + \mathrm{Au}$}

\newcommand{\hehe}[1]{${}^{#1}\mathrm{He} + {}^{#1}\mathrm{He}$}

\newcommand{\he}[1]{\ensuremath{{}^{#1}\mathrm{He}}}

\newcommand{\fakesection}[1]{%
  \par\refstepcounter{section}%
  \sectionmark{#1}%
  \addcontentsline{toc}{section}{\protect\numberline{\thesection}#1}%
  \textbf{#1.}
}

\makeatother

\title{From Lead to Helium: Potential for a Measurement of\\ Jet Quenching in the Smallest Collision Systems}

\author{Coleridge Faraday}
\email{frdcol002@myuct.ac.za}
\affiliation{Department of Physics\char`,{} University of Cape Town\char`,{} Private Bag X3\char`,{} Rondebosch 7701\char`,{} South Africa}

\author{Ben Bert}
\email{brtben004@myuct.ac.za}
\affiliation{Department of Physics\char`,{} University of Cape Town\char`,{} Private Bag X3\char`,{} Rondebosch 7701\char`,{} South Africa}

\author{Jack Brand}
\email{brnjac041@myuct.ac.za}
\affiliation{Department of Physics\char`,{} University of Cape Town\char`,{} Private Bag X3\char`,{} Rondebosch 7701\char`,{} South Africa}

\author{Werner Vogelsang}
\email{werner.vogelsang@uni-tuebingen.de}
\affiliation{Institute for Theoretical Physics\char`,{} University of T\"ubingen\char`,{} Auf der Morgenstelle 14\char`,{} D-72076 T\"ubingen\char`,{} Germany}

\author{W.\ A.\ Horowitz}
\email{wa.horowitz@uct.ac.za}
\affiliation{Department of Physics\char`,{} University of Cape Town\char`,{} Private Bag X3\char`,{} Rondebosch 7701\char`,{} South Africa}
\affiliation{Department of Physics\char`,{} New Mexico State University\char`,{} Las Cruces\char`,{} New Mexico\char`,{} 88003\char`,{} USA}
\affiliation{Theoretical Sciences Visiting Program\char`,{} Okinawa Institute of Science and Technology Graduate University\char`,{} Onna\char`,{} 904-0495\char`,{} Japan}

\date{\today}

\begin{abstract}

We present perturbative quantum chromodynamics (pQCD) predictions for the modification to the yield of high-momentum particles in very light ion collisions---\btbt, \lili, \hehe{4}, and \hehe{3}---both with and without medium-induced energy loss. 
We show that there is non-trivial suppression expected from our partonic energy loss model in symmetric systems from ${}^{208}\mathrm{Pb} + {}^{208}\mathrm{Pb}$ to \hehe{3} and in asymmetric systems $A + B$, and that the energy loss scales approximately with $(\sqrt{A B})^{1 / 3}$. 
Further, we find that deep inelastic scattering measurements in \he{3} and \li{} tightly constrain the nPDF baseline, making these isotopes a particularly clean environment for searching for final-state partonic energy loss induced by the formation of a quark-gluon plasma in these very small systems.

\end{abstract}

\maketitle

\begin{figure}[!b]%
	\centering
	\vspace{-20pt} 
	\includegraphics[width=0.86\linewidth]{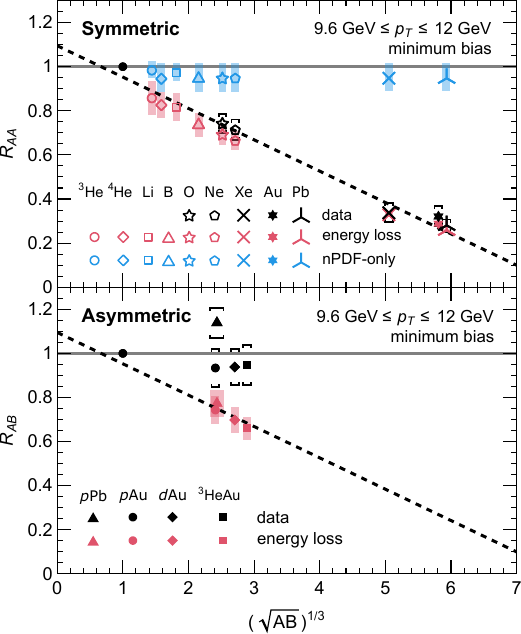}%
	\caption{\small $R_{AB}$ versus $(\sqrt{AB})^{1/3}$ for collisions of nuclei with mass numbers $A$ and $B$. Experimental data for $\pi^0$ and $h^{\pm}$ hadrons produced in \pbpb{} \cite{CMS:2016xef}, \auau{} \cite{PHENIX:2008saf}, \xexe{} \cite{CMS:2018yyx}, \ppb{} \cite{ATLAS:2022kqu}, \pdhau{} \cite{PHENIX:2021dod}, and \oo{} \cite{CMS:2025bta} and \nene{} \cite{2969907,CMS:2026qef} are shown in black. Partonic energy loss predictions (red) and NLO pQCD nPDF-only (blue) are shown for $\pi^0$ hadrons. \oo{}, \nene{}, \pbpb{}, and \xexe{} data were rebinned in \cite{CMS:2026qef,2969907} to $9.6 ~\mathrm{GeV} \leq p_T \leq 12 ~\mathrm{GeV}$, and other data were matched to the closest $p_T$ bins. Systematic theoretical uncertainties are shaded, experimental systematic uncertainties bracketed, and statistical uncertainties are smaller than the markers. Experimental and nPDF uncertainties are shown at $68\%$ CL; energy loss uncertainties represent an envelope of model predictions. A dashed line is fit to symmetric-system data; the solid point at unity denotes $R_{pp} \equiv 1$.}%
	\label{fig:raa_vs_A13}%
\end{figure}

\fakesection{Introduction} Remarkable measurements from the collisions of large nuclei, \auau{} at the Relativistic Heavy Ion Collider (RHIC) and \pbpb{} at the Large Hadron Collider (LHC), show that $\mathcal{O}(10^4)$ particles yield signs of collective behavior in the form of, e.g., near-perfect fluidity \cite{STAR:2000ekf,PHENIX:2003qra,ALICE:2010suc}, strangeness enhancement \cite{STAR:2003jis,ALICE:2013xmt}, quarkonium suppression \cite{STAR:2009irl,ALICE:2012jsl,CMS:2012bms},  and, in particular, the energy loss of high-momentum particles \cite{PHENIX:2001hpc,STAR:2003fka,ALICE:2010yje,ALICE:2012ab,CMS:2012aa}, indicating the production of a novel state of matter, the quark-gluon plasma (QGP). Shocking results from high-multiplicity \ppb{} collisions at LHC \cite{ATLAS:2012cix, ALICE:2012eyl, CMS:2012qk, ALICE:2016sdt, ATLAS:2013jmi, ALICE:2014dwt,CMS:2015yux,CMS:2018loe,ATLAS:2019pvn,ALICE:2024vzv} and \pdhau{} collisions at RHIC \cite{PHENIX:2018lia, STAR:2022pfn} with $\mathcal{O}(10^2)$ particles show many of these same signs of non-trivial emergent many-body behavior consistent with the formation of a QGP in this system.
In the condensed matter context, non-trivial collective motion has been observed in cold fermionic atom systems of $\mathcal{O}(10)$ particles \cite{Brandstetter:2023jsy}.
The minimum bias nuclear modification factor $R^h_{AB}(p_T) \equiv (AB)^{-1} (d \sigma_{\mathrm{AB}}^{h} / d p_T d y)/(d \sigma_{p p}^{h} / d p_T )$ is used to quantify the measurement of the attenuation of high momentum particles, i.e.\ \textit{jet quenching}. 
As shown in \cref{fig:raa_vs_A13} (bottom), the one glaring missing signature for QGP formation in asymmetric small systems in high-energy physics is jet quenching; i.e.\ the measured $R_{AB}$ in asymmetric small systems is inconsistent with energy loss model predictions \cite{ALICE:2014xsp,ATLAS:2022iyq,CMS:2025jbv}.

The substantial theoretical and experimental difficulties involved with making and interpreting energy loss measurements in asymmetric systems---including selection biases in centrality-cut measurements \cite{ALICE:2014xsp}, hard-soft correlations and sensitivity to the internal proton structure arising from the single incident proton \cite{Alvioli:2013vk,  Kordell:2016njg, Perepelitsa:2024eik, JETSCAPE:2024dgu}, the explicit violation of boost-invariance due to the asymmetry, and the absence of well-calibrated 3+1D hydrodynamics models---motivated the collision of symmetric small systems at LHC with a similar multiplicity to central \ppb{} collisions \cite{Citron:2018lsq, Huss:2020dwe,Huss:2020whe, Brewer:2021kiv}.
Recent measurements in \oo{} \cite{ALICE:2026zck,CMS:2025bta} and \nene{} \cite{2969907,CMS:2026qef} collisions indicate non-trivial final-state partonic energy loss of high momentum particles.
This raises the question: what is the smallest possible system in which non-trivial, emergent, many-particle dynamics can be probed by the suppression of high-momentum particles, i.e.\ what is the smallest system in which jet quenching can be measured?

To find evidence for jet quenching, one must be able to distinguish between final-state partonic energy loss and initial-state nuclear parton distribution function (nPDF) effects. Therefore, one needs  well-controlled predictions for final-state partonic energy loss and for initial-state nPDF effects. 
Our perturbative quantum chromodynamics (pQCD)-based energy loss model \cite{Faraday:2025pto} includes elastic and inelastic energy loss channels, small system corrections to these channels, and a rigorous statistical calibration against large system data at RHIC and LHC. Our baseline predictions are from state-of-the-art, next-to-leading-order (NLO) nPDF calculations, rigorously constrained to an unprecedented breadth of data \cite{Eskola:2021nhw}. 
In this work we consider theoretical uncertainties associated with the NLO nPDF production baseline \cite{Huss:2020whe,Huss:2020dwe,Eskola:2021nhw,Borsa:2021ran,Mazeliauskas:2025clt}, theoretical uncertainties associated with the energy loss \cite{Horowitz:2009eb,Faraday:2024gzx,Faraday:2025pto}, and the expected statistical and systematic uncertainties on measurements at LHC \cite{CMS:2025bta,2969907,ALICE:2026zck}. 
In this way we can provide a realistic and quantitative estimate for the distinguishability of the final-state jet quenching signal from the initial-state baseline in small systems.

In \cref{fig:raa_vs_A13} (top), we visualize the distinguishability between final-state energy loss and nPDF-only effects for small systems from \btbt{} to \hehe{3}. 
Also shown in \cref{fig:raa_vs_A13} (top) is the quantitative agreement of our energy loss model with experimental data for symmetric systems from \pbpb{} to \oo{}. 
We find that the isotopes that most separate the energy-loss signal from the initial-state baseline in extremely small systems are \lili{}, which results in an expected significance of $2.3$ ($3.0$) in minimum bias ($0\text{--}10\%$ centrality) collisions, and \hehe{3}, which yields an expected significance of $1.7$ ($2.6$) in minimum bias ($0\text{--}10\%$ centrality) collisions. 
\hehe{3}{} collisions will yield $\sim 30$ particles and \lili{} $\sim 50$, similar to the $\mathcal{O}(10)$ particles of \cite{Brandstetter:2023jsy}.
Surprisingly, \hehe{4} is currently not well suited for measuring jet quenching in small systems.
To make these statements precise, we utilize the following canonical experimental observable.

\fakesection{Nuclear modification factor} The standard candle of final-state partonic energy loss in high-energy hadronic collisions is the nuclear modification factor,
\begin{equation}
	R^h_{AB}(p_T) = \frac{1}{N_{\text{coll}}} \frac{dN^h_{AB} / d p_T}{dN^h_{pp} / d p_T},
	\label{eqn:}
\end{equation}
where $dN^h_{AB} / dp_T$ and $dN^h_{pp} / d p_T$ are the $p_T$-differential numbers of hadrons $h$ produced in $A + B$ and \pp{} collisions, respectively, and $N_{\text{coll}}$ is the number of binary collisions, typically computed by experiments using the Glauber model \cite{Glauber:1970jm,Miller:2007ri}. 
In heavy-ion collisions, the path-length dependence of energy loss is commonly studied by selecting events in different centrality classes, which correlate with the impact parameter and thus with the size of the nuclear overlap region and the typical path length traversed by a hard probe.
In smaller systems, however, the correlation between centrality measures and impact parameter is significantly reduced and there are selection biases as well as substantial model uncertainties associated with performing centrality cuts \cite{ALICE:2014xsp}. Therefore, as was suggested for \oo{} collisions \cite{Huss:2020dwe,Huss:2020whe}, and then subsequently measured \cite{CMS:2025bta}, the minimum bias nuclear modification factor is a good candidate for measuring energy loss in such small systems,
\begin{equation}
	R^h_{AB} = \frac{1}{AB} \frac{d\sigma^h_{AB} / d p_T}{d\sigma^h_{pp} / d p_T}.
	\label{eqn:min_bias_raa}
\end{equation}

Despite the greater difficulties in measuring and interpreting centrality-cut observables, we nevertheless report the $0\text{--}10\%$ most central collision suppression predictions from our energy loss model. 
We are motivated to report these results as our model predicts considerably more suppression in the $0\text{--}10\%$ centrality class than in minimum bias collisions, which increases the expected significance of any measurement of energy loss.

\fakesection{Baseline perturbative QCD expectation} An important aspect of interpreting any potential measurement of the $R_{AB}$ as partonic energy loss is a comparison to the baseline perturbative QCD expectation. That is, the prediction for the $R_{AB}$ assuming that the initial-state spectrum of partons is modified in the nuclear collision only by changes to the parton distribution functions and not, e.g., by final-state energy loss. Collinear factorization is a framework in QCD that is typically assumed for all high momentum transfer processes and is proven for a subset of these processes \cite{Collins:1989gx}. Within collinear factorization, the cross section for the production of a hadron $h$ in a collision of two nuclei $A$, $B$ can be written as
\begin{align}
	&d\sigma_{AB\to hX} = \sum_{a,b,c} f_{a/A}(x_a,\mu_F) \otimes f_{b/B}(x_b,\mu_F)\label{eqn:factorization_nn}\\
	&\otimes d\hat{\sigma}_{ab}^c(x_aP_A,x_b P_B,P_h/z,\mu_R,\mu_F,\mu_F') \otimes D_{c \to h}(z,\mu_F'),\nonumber
\end{align}
where $P_A,P_B$ and $P_h$ denote the momenta of the two nuclei and the observed hadron, respectively; 
$f_{a/A}$ and $f_{b/B}$ are the nPDFs; and $D_{c \to h}$ is the fragmentation function (FF) for parton $c$ giving rise to the observed hadron $h$. Finally, the $d\hat{\sigma}_{ab}^c$ are the partonic hard-scattering cross sections for initial partons $a,b$ producing the outgoing parton $c$. These cross sections are amenable to QCD perturbation theory. The various functions in \cref{eqn:factorization_nn} are tied together by integrations over the partons' momentum fractions, as denoted by the $\otimes$ symbol. They are also linked by their dependence on renormalization ($\mu_R$) and factorization ($\mu_F,\mu_F'$) scales, the latter distinguished by whether they are associated with the initial or final parton state. 

The nPDF-only baseline nuclear modification factor is then constructed as the ratio of the $AB$ cross section to the $pp$ one:
\begin{equation}
	R_{AB}^{h, \text{nPDF}}(p_T, y) = \frac{1}{A B} \frac{d \sigma_{AB}^{h + X} /d p_T dy}{d \sigma_{pp}^{h + X} / d p_T dy},
	\label{eqn:raa_npdf_nn}
\end{equation}
where the cross sections are typically differential in $p_T$ and $y$. In this work we take $|y| < 1$, and compute $R_{AB}$ at next-to-leading order, using the program developed in \cite{Jager:2002xm}. For the nPDFs, we adopt the EPPS21~\cite{Buckley:2014ana,Eskola:2021nhw} sets, along with their CT18A proton baseline PDFs~\cite{Hou:2019efy}. 
EPPS21 has larger uncertainties and more suppression than other popular nPDF sets \cite{Mazeliauskas:2025clt}, for example, TUJU21 \cite{Helenius:2021tof}, nNNPDF3.0 \cite{AbdulKhalek:2022fyi}, and nCTEQ15 \cite{Kusina:2015vfa}. Therefore, our choice to use EPPS21 is the most conservative for estimating the expected significance of an energy-loss signal in light-ion collisions.
For the FFs we use the recent BDSS set for $\pi^0$ of \cite{Borsa:2021ran} which includes information from $pp$ scattering at RHIC and the LHC, as well as from semi-inclusive deep inelastic scattering (DIS), and hence has better constrained gluon and flavor-separated quark fragmentation functions than previous sets. Generally, as already observed in \cite{Huss:2020dwe,Huss:2020whe}, the precise choice of fragmentation functions is not very important as any changes of the FFs largely cancel in the ratio $R_{AA}^{h, \text{nPDF}}$. %

The primary uncertainties of the predictions for $R_{AA}^{h, \text{nPDF}}$ result from those of the nuclear PDFs. These uncertainties may be readily assessed by using the ``error sets'' of PDFs also provided within EPPS21 for each nucleus. The (smaller) uncertainties associated with the proton PDFs are reflected in these error sets as well. We follow the technique described in~\cite{Eskola:2021nhw} to compute $68\%$ confidence level (CL) PDF uncertainties of our $R_{AA}^{h, \text{nPDF}}$. A further source of uncertainty comes from the dependence on the scales $\mu_R,\mu_F$, and $\mu_F'$. We use $\mu_R=\mu_F=\mu_F'=p_T$ as our baseline choice, but vary each scale separately by a factor of two up and down from this default scale. We discard, however, very ``asymmetric'' scale choices for which one of the scales is $p_T/2$ while another is $2p_T$. In this way, we end up with a ``15-point'' scale variation scheme. We always vary the scales in the same way in the numerator and in the denominator of $R_{AA}^{h, \text{nPDF}}$ since this seems most appropriate physically. We then generally find that the scale dependence cancels to a large extent in $R_{AA}^{h, \text{nPDF}}$, even though the individual cross sections are quite sensitive to the scale choice.

\fakesection{Partonic energy loss predictions} There is a large literature of partonic energy loss calculations in various frameworks that make different assumptions \cite{Baier:1996sk, Baier:1996kr, Zakharov:1996fv, Zakharov:1997uu, Wiedemann:2000za, Gyulassy:2000er, Wang:2001ifa} as well as many phenomenological implementations of these energy loss calculations that have been applied to small systems \cite{JETSCAPE:2024dgu,Huss:2020dwe,Huss:2020whe,Ke:2022gkq,Zakharov:2021uza,vanderSchee:2025hoe,Pablos:2025cli,Zakharov:2025mbk}. We use our pQCD-based energy loss model that is described in detail in our previous work \cite{Faraday:2025pto}, briefly recapping the main features here.

\begin{figure*}[t]
  \includegraphics[width=\linewidth]{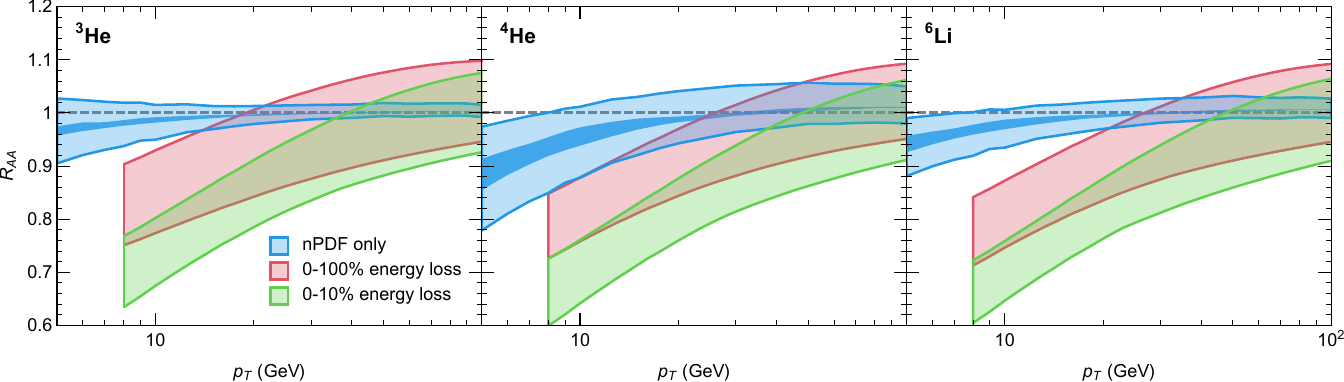}
  \caption{Plot of $R_{AA}$ as a function of $p_T$ for $\pi^0$ hadrons produced in \hehe{3}{} (left), \hehe{4}{} (central), and \lili{} (right) collisions. Predictions from our partonic energy loss model are shown for $0\text{--}100\%$ centrality (red) as well as $0\text{--}10\%$ centrality (green), where the band represents the envelope of all energy loss results computed with different variations in the theoretical uncertainty. The nPDF-only predictions are also shown (blue), where the lighter, outer band represents the nPDF and PDF uncertainties and the darker, inner band represents the scale uncertainties. }
  \label{fig:raa_pt_he3}
\end{figure*}

Partons propagate through a fluctuating collision geometry \cite{Schenke:2020mbo} that is generated with IP-Glasma initial conditions \cite{Schenke:2012wb, Schenke:2012hg} and evolved according to longitudinal Bjorken expansion \cite{Bjorken:1982qr}. Initial nucleon locations are sampled event-by-event using Woods-Saxon distributions for \pb{}, \au{}, and \xe{} with standard parameters \cite{Schenke:2020mbo}; Projected Generator Coordinate Method (PGCM) \cite{Frosini:2021sxj,Frosini:2021fjf,Frosini:2021ddm} sampling for \neon{} and \ox{}, nucleon configuration for which were obtained from the IP-Glasma code \cite{Schenke:ipglasma,Mantysaari:2025tcg}; and a variational Monte Carlo (VMC) for \he{3} \cite{Nagle:2013lja} and \he{4}{} \cite{Carlson:2014vla,Lim:2018huo}. While both \li{} and \bt{} are small nuclei, and so should in principle be modeled similarly to \he{} or \neon{} / \ox{}, in this first work we model them with a Woods-Saxon distribution as only the one- and two-nucleon distributions from these more sophisticated nucleon configurations are available. The Woods-Saxon parameters are found to be $R = 1.355 ~\mathrm{fm}$ and $a = 0.584 ~\mathrm{fm}$ for \li, and $R = 2.367 ~\mathrm{fm}$ and $a = 0.349 ~\mathrm{fm}$ for \bt{} by fitting to single nucleon densities computed in the same VMC framework as \he{3} and \he{4} \cite{Wiringa:2013ala,Wiringa:2025}. We checked that modeling \li{} (\bt{}) as a bound state of an alpha particle (two alpha particles) and a deuteron only increased the predicted $R_{AA}$ by $\lesssim 1\%$. Similarly, including a large deformation parameter $\beta_2 =0.8$ for \bt{} increased the predicted $R_{AA}$ by $\sim 1\%$. A minimum nucleon separation of $0.9 ~\mathrm{fm}$ is enforced for all Woods-Saxon implementations \cite{Schenke:2020mbo}. 

We note that \li{} and \bt{} contain both odd neutron and proton number, and are therefore subject to larger theoretical uncertainties in nuclear structure calculations compared to even-even nuclei \cite{Carlson:2014vla,Myo:2015oia}. However, such uncertainties are most relevant for observables sensitive to the detailed shape and fluctuations of the nuclear geometry, such as $v_n$ and correlation observables. As evidenced by our insensitivity to a wide variety of assumptions about the initial nucleon configuration, we anticipate that our energy loss predictions will not be significantly affected by these uncertainties.

Parton production is assumed to be the same in \pp{} as in \aaa{}. Including nPDF effects would likely be largely absorbed in a rigorous statistical recalibration to large system data. As one moves to smaller systems, where nPDF effects constitute a larger fractional contribution, the predicted suppression from our energy loss model would presumably increase. Our conclusions are therefore conservative with respect to the magnitude of suppression due to energy loss in small systems.

We compute both collisional and radiative energy loss using pQCD-based calculations. Radiative energy loss is computed in the DGLV framework \cite{Djordjevic:2003zk} including corrections that are exponentially suppressed by path length \cite{Kolbe:2015rvk} and neglected by all other models \cite{Faraday:2023mmx}, making our calculation particularly well-suited to small systems. Collisional energy loss is computed using two limiting cases: hard thermal loops (HTL)-only \cite{Wicks:2008zz} and Braaten-Thoma (BT) \cite{Braaten:1991we}. Collisional energy loss is essential for reliable small system predictions, as it dominates radiative loss ($L$ vs $L^2$ scaling) for very short paths \cite{Faraday:2024gzx}.

We treat the effective strong coupling $\alpha_s^{\text{eff.}}$ as our single free parameter, neglecting running coupling effects. We previously estimated the uncertainty from this approximation by allowing the coupling to run over commonly used phenomenological scales and found the uncertainty to be small when extrapolating our results to small systems \cite{Faraday:2024qtl}.
The effective strong coupling is fit to all available high-$p_T$ $R_{AA}$ data from charged hadrons, pions, and $D$ mesons with $8 ~\mathrm{GeV} \leq p_T \leq 50 ~\mathrm{GeV}$ in $0\text{--}50\%$ centrality \pbpb{} collisions \cite{Faraday:2025pto}. Theoretical uncertainties are estimated from two sources: varying the collinear radiation cone by factors of two up and down \cite{Horowitz:2009eb,Faraday:2025pto}, and the choice between the two collisional kernels, HTL-only and BT, which treat the HTL-vacuum propagator transition in limiting ways. The fit is performed separately for each model variation, and the envelope of results defines our theoretical uncertainty band.	We find the best fit extracted is $\alpha_s^{\text{eff.}} = \num{0.37(11:8)}$ at LHC \cite{Faraday:2025pto}. While it is difficult to translate our extracted $\alpha_s^{\text{eff.}}$ to the jet transport coefficient, our best translation gives a jet transport coefficient normalized to temperature cubed of $\hat{q} / T^3 = 11\pm 3$ \cite{Faraday:2025pto}. 

Other important theoretical uncertainties include energy loss pre-thermalization, running coupling effects, minimum bias trigger inefficiency, and the sensitivity to the nuclear geometry. While we do not include these uncertainties in \cref{fig:raa_vs_A13}, we quantitatively estimated their effects for \hehe{3}{} and \lili{} and found that they only increased the expected suppression of our results. Therefore we anticipate that our results are conservative with respect to the expected suppression in small systems. %

\fakesection{Results} \Cref{fig:raa_pt_he3} plots $R^{\pi^0}_{AA}$ from nPDFs only and from our energy loss model as a function of $p_T$ for \hehe{3}, \hehe{4}, and \lili{} collisions, respectively.
Comparing these figures, we see that the nPDF-only prediction for \hehe{4} has both the central value lower and the uncertainty band larger than that of \hehe{3} and \lili. This reflects the specialized parameterization for \he{3}{} and \li{} in EPPS21 \cite{Eskola:2021nhw}, which was necessary to describe DIS data and is physically motivated by the tighter nuclear binding of \he{4} compared to that of \he{3} and \li{} \cite{Wang:2012eof}. In minimum-bias collisions, where selection biases and hard-soft correlations are minimized, \lili{} shows the clearest separation between predictions, with distinct differences in both magnitude and $p_T$ dependence near $p_T \simeq 10 ~\mathrm{GeV}$.

To quantify experimental sensitivity, we estimate whether a measurement consistent with energy loss could reject the nPDF-only null hypothesis. 
We assume systematic uncertainties dominate and are similar in magnitude to \neon{} measurements ($\sim 5\%$, combining normalization and systematics in quadrature). Statistical uncertainties for the $R_{AA}$ in \nene{} collisions were negligible up to $p_T \simeq 50 ~\mathrm{GeV}$, even with only a single fill of data taking \cite{2969907,CMS:2026qef}. 
While assigning probability distributions to theoretical uncertainties is inherently difficult, we proceed by treating our energy loss envelope as covering $90\%$ of physically reasonable pQCD-based models. %
Under these assumptions, the expected significance is $N_\sigma = | \mu_{\text{nPDF}} - \mu_{\text{EL}}|/(\sigma_{\text{nPDF}}^2 + \sigma_{EL}^2 + \sigma_{\text{exp}}^2)^{1 / 2}$, where $\mu$ denotes central values and $\sigma$ denotes uncertainties. Evaluating this for $9.6 ~\mathrm{GeV} \leq p_T \leq 12 ~\mathrm{GeV}$ $\pi^0$ hadrons (see \cref{fig:raa_vs_A13}), we find $N_{\sigma} = 1.7, \; 1.5, \; 2.3,$ and $2.7$ for \hehe{3}, \hehe{4}, \lili{}, and \btbt{}, respectively. For comparison, we find $N_{\sigma} = 3.4$ for \oo. 
Remarkably, lithium achieves $\sim 70\%$ of the oxygen sensitivity despite having less than half the nucleons, offering a unique opportunity to search for final-state partonic energy loss due to quark-gluon plasma formation in a system significantly smaller than oxygen-oxygen collisions.

If centrality biases can be overcome, for example by normalizing with electroweak bosons \cite{PHENIX:2023dxl}, spectator nucleons at zero degrees \cite{ALICE:2014xsp}, or self-normalizing dijet probes \cite{ATLAS:2022iyq, CMS:2025jbv}, then $0$--$10\%$ centrality-selected collisions would provide significantly larger suppression than minimum bias. We note that for the lightest systems considered here, the small neutron number limits ZDC-based methods to only a few coarse centrality bins.
If we conservatively double the experimental uncertainty to $10\%$ to account for the more complicated measurement of the centrality-cut observable, we find $N_{\sigma} = 2.6, \; 2.3, \; 3, $ and $3.5$ for \hehe{3}, \hehe{4}, \lili, and \btbt, respectively. Assuming one can overcome centrality biases, we see that any of these systems would be able to disentangle initial- from final-state effects at $\gtrsim 2 \sigma$ level; however, \hehe{3} would be particularly compelling given the extremely small size of the formed plasma.

There are of course many isotopes smaller than ${}^{16}\mathrm{O}$ that we could have considered. We have focused here on isotopes with $Z / A \geq 1 / 2$, those significantly smaller than ${}^{16} \mathrm{O}$, and those that are feasible for collisions at the LHC. \btbt{} is particularly interesting because, even though it is not particularly smaller than ${}^{16} \mathrm{O} + {}^{16} \mathrm{O}$, this isotope is planned to be developed for the SPS fixed target physics in Run 4 and could potentially be ``parasitically" sent to the LHC.
The measurement of $R_{AA}$ in \btbt{} collisions would provide a non-trivial extrapolation down in $A$ and also a multiplicity similar to minimum bias \ppb{} collisions ($\sim 100$ for \btbt{} compared to $\sim 75$ for \ppb).
We did not consider deuterium because there are some non-trivial considerations for the nPDFs of deuterium \cite{Eskola:2021nhw} and we expect that the energy loss would be so small such that it would be difficult to distinguish from baseline predictions.

\fakesection{Conclusions} We have shown that our pQCD-based energy loss model leads to an approximate linear dependence of the nuclear modification factor on the cube root of the geometric average of the atomic mass numbers involved in the collision, i.e.\ $R_{AB} \propto (\sqrt{AB})^{1 / 3}$. We see good experimental agreement with this trend from the measured $R_{AA}$ for symmetric \pbpb, \xexe, \nene, and \oo; however, we see a striking disagreement with the measured $R_{AB} \simeq 1$  in minimum bias asymmetric \pdha{} collisions compared to our theoretical predictions of $R_{AB} \simeq 0.7\text{--}0.8$. We argued that this disagreement shows that there is some physics---either experimentally, theoretically, or both---that is not understood in asymmetric systems. We note that the linear dependence of the $R_{AA}$ on $A^{1/3}$ is emergent and a product of many interplaying effects---the $L$ ($L^2$) dependence of collisional (radiative) energy loss, the time dependence of the temperature, the convolution with the steeply falling production spectrum, and the realistic distribution of binary collisions---and therefore should not be interpreted as evidence for energy loss scaling with $L$.

To avoid the experimental and theoretical difficulties associated with asymmetric systems, and inspired by the recent observation of high-$p_T$ hadron suppression in ${}^{16}\mathrm{O} + {}^{16}\mathrm{O}$ collisions \cite{CMS:2025bta}, we investigated the possibility of measuring energy loss in the smallest possible symmetric hadronic collision systems.
We studied a variety of systems with atomic mass number $A$ less than that of oxygen, including \he{3}, \he{4}, \li, and \bt, by computing the expected $R_{AA}$ with and without medium-induced, partonic energy loss. 
We showed that, in minimum-bias collisions, \lili{} occupies a unique sweet spot: a very small system with fewer than half the nucleons of oxygen, yet offering comparable sensitivity for distinguishing energy-loss effects from nPDF-only scenarios.
If one could overcome the challenges associated with centrality-cut measurements in small systems, the small uncertainty in the \hehe{3} nPDF-only baseline and the energy loss predicted from our calculations implies that a $\sim 2.5 \sigma$ energy-loss signal could be measured in this tiny system. 
Ongoing analyses of $p + \mathrm{O}$ collisions will further constrain the nPDF predictions for \hehe{3}{} and \lili{} $R_{AA}$ and thus will more cleanly separate the baseline from the signal. 
It has also been proposed to consider the double ratio $R_{AA} / R_{pA}^2$, in which correlated nPDF uncertainties cancel to within $1\%$ \cite{Jonas:2026yoz}. Such a measurement would require an additional $p + A$ run, but would increase the expected significances to $2.2,~3.1,~3.2$ and $5.3 \sigma$ for \hehe{3}, \hehe{4}, \lili{}, and \btbt{}, respectively.

Future collisions at the LHC that involve \li{} and \he{3}{} offer a tantalizing opportunity to test the emergence of collective behavior due to the formation of a new state of matter in systems of $\mathcal{O}(10)$ particles, extending the recent observations from electromagnetic interactions at $0 ~\mathrm{K}$ in \li{} atoms \cite{Brandstetter:2023jsy} to strong interactions at $10^{12}~ \mathrm{K}$ in \li{} and \he{3}{} nuclei.

\fakesection{Acknowledgments} The authors thank the organizers and participants of the \textit{Light Ion Collisions at the LHC 2025} workshop, which was formative in generating the ideas in this manuscript. Conversations with Nico Orce, Giuliano Giacalone, Jamie Nagle, Petja Paakkinen, Hannu Paukkunen, John Jowett, Maciej Slupecki, Florian Jonas, and Florian Lindenbauer were particularly insightful. We thank Cristian Baldenegro for providing us with preliminary CMS data. CF and WAH thank CERN-TH for hospitality during the course of this work. 
This research was conducted in part by WAH while visiting the Okinawa Institute of Science and Technology (OIST) through the Theoretical Sciences Visiting Program (TSVP). 
Computations were performed using facilities provided by the University of Cape Town’s ICTS High Performance Computing team: \href{http://hpc.uct.ac.za}{hpc.uct.ac.za}. This work was supported in part by the National Research Foundation of South Africa (reference numbers PMDS240606224849 and 65212). CF, BB, and WAH thank the National Research Foundation and the SA-CERN collaboration for their generous financial support during the course of this work. CF thanks the National Institute for Theoretical and Computational Sciences (NITheCS) and the University of Cape Town Science Faculty PhD Fellowship for their generous financial support.
WV is grateful to the Federal Ministry of Education and Research (BMFTR), grant no.\ 05P21VTCAA.

\fakesection{Author Contributions} WAH and CF jointly conceptualized the study. CF performed the energy-loss and hydrodynamic calculations and wrote all drafts. BB and JB performed exploratory PYTHIA simulations and cross-checked the nPDF uncertainty implementation. WV performed the NLO pQCD calculations with nPDFs and wrote most of the section on nPDF results. WAH supervised the project. All authors edited the manuscript and contributed to the interpretation and scientific conclusions.

\balance
\bibliographystyle{apsrev4-2} %
\bibliography{manual,HeHeSuppression.bib}

\end{document}